\newcommand{\ee}{\end{equation}}

\documentstyle[psfig,11pt,fullpage]{article}
\def\boxit#1{\vbox{\hrule\hrule\hrule\hbox{\vrule\vrule\vrule\kern5pt\vbox{\kern5pt#1\kern5pt}\kern5pt\vrule\vrule\vrule}\hrule\hrule\hrule}}
\def\comma{ \; , }
\def\period{ \; . }

\begin{document}
\onecolumn

\begin{center}

{\Large {\bf {Is mass conformally invariant?}}} \\ 
\vspace{0.2cm}
Kevin C.K. Chan \\
\vspace{0.2cm}
Department of Physics, University of Waterloo, Waterloo, Ontario, Canada N2L 3G1 \\
kckchan@avatar.uwaterloo.ca

\end{center}

\centerline{ABSTRACT}

\vspace{0.1cm}

By using the Garfinkle, Horowitz and  Strominger
black hole solutions as examples,  we illustrate that,  with respect to the 
reference action functional  proposed by Hawking and Horowitz, 
the asymptotic mass parameter is not invariant between 
two conformally related  static spherically symmetric metrics.  

\section{Introduction}

Given two conformally related static spherically symmetric (SSS) metrics,  
it is generally believed that the conformal transformation does not change
a physical quantity such as mass \cite{horowitz}. In a previous paper we briefly pointed out that the situation is somewhat more subtle \cite{ccm}. In this paper, we discuss this issue more explicitly.  We illustrate the conformal non-invariance of quasilocal and asymptotic mass\footnote{The word ``quasilocal'' refers to a compact orientable spatial two-surface, usually spherical. Asymptotic mass is the quasilocal mass evaluated at spatial infinity. In general relativity, the asymptotic mass in an asymptotically flat spacetime is called the ADM mass.} by using the the Hawking and Horowitz (HH) prescription \cite{hh}  and the  Garfinkle, Horowitz and Strominger (GHS) string black hole solutions obtained in \cite{ghs}. We will calculate (i) the asymptotic mass directly from the string action and metrics for the GHS black holes, and (ii) the asymptotic mass directly from the conformally related Einstein action and metric. Then we show that the asymptotic mass in the string metr!
ics is different from the one in t
he Einstein metric.  As an aside, we will discuss an improper method  that was used in \cite{xz} to  identify the asymptotic mass parameter in scalar-tensor-typed gravity.  Instead of repeating the derivation of formulae and material that had readily given in \cite{ccm, hh, broyor}, we simply quote the results, apply them to the GHS solutions, and then discuss the significance.  We also restrict our discussions to only SSS spacetimes. 

\section{HH prescription}

Consider a scalar-tensor gravity of the following form, whose metrics are
given by $g_{\mu\nu}$. 
\begin{equation}
S = \int_{\cal M} d^nx\sqrt{-g}[D(\phi){\cal R}+H(\phi)(\nabla\phi)^2+V(\phi)]
+S_{matter}+S_{boundary}+S_0  \period \label{E1}
\end{equation}
$D(\phi)=1$ corresponds to general relativity minimally couples to a scalar (dilaton). 
$S_0$ is the background action which specifies a reference 
spacetime that effectively define zero mass.  
It is a functional of the fields on the boundary, 
$\partial{\cal M}$, of the spacetime. 
Consider the case of  a static spherically 
symmetric (SSS) solution to the  field equations derived from 
(\ref{E1}). 
The SSS metric can be written in the following form
\begin{equation}
ds^2 = -N^2(r)dt^2+\frac{dr^2}{f^2(r)}+r^2d\Omega^2 \period 
\label{E2}
\end{equation}
The HH prescription can be briefly summarized as follows \cite{ccm, hh}: $S_0$ has the same functional form as the rest of $S$ and is a functional of the variables $(g_0)_{\mu\nu}$ and $\phi_0$ which are independent of  $g_{\mu\nu}$ and $\phi$ except on the  boundary of the spacetime where $N_0=N$ and $\phi_0=\phi$ are required. The reference spacetime is a solution to the field equations produced from $S_0$. Under this HH prescription, the quasilocal  mass $M(r)$ of SSS metric (\ref{E2}) is given by \cite{ccm, hh}
\begin{equation}
M(r)  = N\left[\frac{r^2}{2}\left(D^{'}+\frac{2D}{r}\right)(f_{0}-f)\right] \comma \label{mass}
\end{equation}
where the prime denotes the derivative with respect to $r$.  
$M(r)$ is the conserved quasilocal charge associated with the timelike
killing vector of the spacetime time,  $(\partial/\partial t)^a$.
This property is not shared by the quasilocal energy $E(r)$ which is given by
\begin{equation}
E(r) = \frac{r^2}{2}\left(D^{'}+\frac{2D}{r}\right)f-E_0 \label{energy}
\end{equation}
It is $E(r)$, not $M(r)$, that is the thermodynamic internal energy which appears in the first law. This point will be significant when we discuss Hawking temperature later. Since the thermodynamic relation is independent of the choice of the background action, we leave $E_0$ as arbitrary.  The asymptotic mass parameter at spatial infinity is defined as $M=M(\infty)$.

\section{GHS black holes and their mass}

We quickly review the GHS black hole solutions. The low energy string action is 
\begin{equation}
S=\int d^4x\sqrt{-g}e^{-2\phi}({\cal R} + 4(\nabla\phi)^2-F^2) \comma \label{saction}
\end{equation}
where $F^2$ is the usual Maxwell contribution. Now we have 
$D(\phi)=e^{-2\phi}$. The magnetic string black hole solution to 
the field equations obtained from (\ref{saction}) is given by 
\cite{ghs}
\begin{eqnarray}
ds^2  (string)  & = &
-\frac{\left(1-\frac{2C}{r}\right)}{\left(1-\frac{Q_m^2 }
{Cr}\right)}dt^2 + \frac{dr^2}
{\left(1-\frac{2C}{r}\right)\left(1-\frac{Q_m^2}{Cr}\right)}
+r^2d\Omega^2 \comma \label{msbh}                                                                       
\end{eqnarray}
\begin{equation}
e^{-2\phi}=1-\frac{Q_m^2}{Cr}  \quad {\rm and} \quad   
F_{\theta\varphi}= Q_m sin\theta   
\period \label{mdil}
\end{equation}
$Q_m>0$ is the magnitude of the magnetic charge and $C$ is a 
constant of integration. We set the asymptotical value of $\phi$, $\phi_0$, to be zero
for simplicity. 
The electrically charged solution is \cite{horowitz}
\begin{eqnarray}
ds^2 (string) & = &
-\frac{\left(1+\frac{Q_e^2-2C^2}{Cr}\right)}
{\left(1+\frac{Q_e^2}{Cr}\right)^{2}}dt^2 +
\frac{dr^2}{\left(1+\frac{Q_e^2-2C^2}{Cr}\right)} +r^2d\Omega^2 \comma \label{esbh}
\end{eqnarray}
\begin{equation}
e^{-2\phi}=1+\frac{Q_e^2}{Cr} \quad {\&} \quad   
F_{tr} =\frac{Q_ee^{4\phi}}{r^2} 
\period \label{edil}
\end{equation}
$Q_e>0$ is the magnitude of the electric charge. 

Using the radial co-ordinate transformations $\rho =r$ 
and $\rho=r+\frac{Q_e^2}{C}$ in the magnetic and electric cases 
respectively, and the following conformal transformation 
\begin{equation}
g^S_{\mu\nu} =e^{2\phi}g^E_{\mu\nu}\comma \label{ct}
\end{equation}  
with (\ref{mdil}) and (\ref{edil}), it is easy to see that the 
corresponding Einstein metric is given by
\begin{eqnarray}
ds^2  (Einstein) & = & 
-\left(1-\frac{2C}{\rho}\right)dt^2+\frac{d\rho^2}
{1-\frac{2C}{\rho}}+\rho\left(\rho-\frac{Q^2}{C}\right)d\Omega^2 
\comma \label{einstein}
\end{eqnarray}
\begin{equation}
e^{\pm 2\phi}=1-\frac{Q^2}{C\rho} \period \label{emdil}
\end{equation}
The corresponding Einstein action is 
\begin{equation}
S=\int d^4x\sqrt{-g}({\cal R} - 2(\nabla\phi)^2- e^{-2\phi}F^2) \comma \label{eaction}
\end{equation}
The plus (minus) sign in (\ref{emdil}) corresponds to the electric 
(magnetic) case. $Q$ denotes the magnitude of  either magnetic or electric charge.  

Now we illustrate the conformal non-invariance of asymptotic mass in the GHS  black holes by using (\ref{mass}) to identify the asymptotic mass parameter in the magnetic and electric black holes in the string and Einstein frames respectively. 
In each case, the background metric can be simply achieved by setting constants of 
integration of a particular solution to some special value 
that then specifies the reference. Typically, Minkowski spacetime will be a SSS solution to the field equations generated by the reference action functional;  
if this is chosen as the reference spacetime, then we have $C_0=Q_{m0}=Q_{e0}=0$  ($f_0=1$ and $\phi_0=0$).    
For the magnetic black hole, we get
\begin{equation}
M_S  =  C + \frac{Q_m^2}{2C}  \period \label{madm} 
\end{equation}
For the electric case (\ref{esbh}), one has 
\begin{equation}
M_S = C -\frac{Q_e^2}{2C} \period \label{eadm}
\end{equation}
By using (\ref{mass}) with $D(\phi)=1$ in the action (\ref{eaction}), Einstein metric  (\ref{einstein}), the corresponding dilaton (\ref{emdil}), and the Minkowski metric and vanishing dilaton as the background solution again, we get
\begin{equation}
C=M_E \period  \label{einadm1}
\end{equation} 
$M_E$ is also called the ADM mass in Einstein frame. 
The result can be summarized as follows:
\begin{equation}
M_S=M_E\pm \frac{Q^2}{2M_E}, \label{result}
\end{equation}
where the plus (minus) sign corresponds to magnetic (electric) solution.
Now one can see that the asymptotic mass parameters in string and Einstein 
cases are different under the HH prescription. 
Note that this conformal non-invariance not only happens at spatial infinity, but also happens quasilocally \cite{ccm}.  For a given positive $M_E$ in the Einstein frame, we see that the mass of magnetic (electric) black hole in the string frame is greater (less) than $M_E$.  This seems impossible since a given black hole should ``weigh'' the same in any frame. The conformal transformation  (\ref{ct}) is simply a field redefinition: define a new metric tensor from an old one. Thus all the physical quantities should be invariant. We will return to this ``puzzle'' in the next section. 

There is one special situation where mass is conformally invariant in the HH prescription.  Assume that the conformal transformation (\ref{ct}) (or its generalization to $D^{-1}$) does not depend on  integration constants which relate to the asymptotic mass parameter. Now using  (\ref{mass}) it is easy to show that quasilocal mass is invariant, regardless of the asymptotic condition of the spacetime. Examples include  a class of asymptotically non-flat black hole solutions to (\ref{saction}) where $e^{2\phi}$ does not depend on any constant of integration \cite{CHM}.  

From now on, we will denote $C$ as $M_E$  using (\ref{einadm1}). For the magnetic case (\ref{msbh}), $M_E$ must always be positive for a positive mass in (\ref{madm}).  The extreme limit is  $\sqrt{2}M_E = Q_m$ or $M_S=\sqrt{2}Q_m$.
For the electric case (\ref{esbh}),  if one rejects the notion of negative values of $M_S$ and $M_E$,  (\ref{eadm}) indicates that one must always have  $2M_E^2\geq Q_e^2 $.
In the extreme limit, $2M_E^2=Q_e^2$. Thus the extreme electric black hole is given by 
\begin{equation}
ds^2=-\left(1+\frac{2M_E}{r}\right)^{-2}dt^2+dr^2+r^2d\Omega^2 \comma \quad 
2M_E^2 = Q_e^2 \period  \label{extreme}
\end{equation}
There is an interesting feature of this solution: It has zero mass quasilocally and asymptotically since $f_0=f=1$ in (\ref{mass}) and consequently $M(r)=0$ for all relevant values of $r$.  
In addition, this extreme solution is horizon-free and has a  curvature singularity at (i) $r=0$ if $M_E>0$ or (ii) $r=2M_E$ if $M_E<0$.  This form of the metric was previously discussed in \cite{horowitz, HT} and $M_E>0$ was considered by the authors as the asymptotic mass in the above string metric.  

\section{Conformally invariant background?}

Obviously, mass is invariant under a co-ordinate transformation. 
Mass must also be invariant under a field redefinition. However, 
a conformal transformation like (\ref{ct}) is a very special field redefinition in two senses. First, in the Einstein frame, the dilaton can be considered as ``pure matter'' in the energy-momentum tensor. It can be separated from gravity. In the string frame, however, the dilaton and the metric tensor together yield the effect of  gravity. Thus a conformal transformation changes the role of dilaton. 
Second, the conformal transformation, $e^{2\phi}$, may contain integration constants which have physical significance (i.e. they cannot be rescaled away). Intuitively, when going from an old metric tensor to a new metric tensor, $e^{2\phi}$ may bring these integration constants to the old metric tensor to change some of its physical quantities. Thus we should treat conformal transformations with care. 

In fact, when calculating the quasilocal and asymptotic mass, we found in \cite{ccm} that one must have a conformally form invariant background action to yield an invariant mass. Thus conformal invariance in mass is not automatically guaranteed for any choice of background action. Obviously, the background action under the HH prescription is not conformally invariant. In fact, every quantity such as mass or internal energy which depends on the choice of the background may suffer similar conformal non-invariance. On the other hand, the definitions of Hawking temperature and entropy are background independent thus they are conformally invariant \cite{ccm, ted}. Similarly we expect that the laws of physics are background independent. For example, the first law of black hole thermodynamics is also valid in string frame \cite{ccm}. In that sense, conformal transformations produce no new laws of physics. 
However, it can change physical quantities. 

Can we find a conformally invariant $S_0$ such that it yields a conformally invariant mass?  Since there is  freedom to choose $S_0$, it is interesting to find  what $S_0$ will yield a conformally invariant mass.  The first candidate is $S_0=0$ but it yields a diverging asymptotic energy and mass for any solution. For a non-vanishing $S_0$,  one first demands conformal invariance and then looks for the required $S_0$.  We found in \cite{ccm} that, for a SSS metric, there is an unique choice of $S_0$ which yields a conformal invariant mass:
\begin{equation}
S_0=-2\int_{\partial {\cal M}} {\bf {\epsilon}} N  D(\phi) k, \label {CI}
\end{equation}
where  $k$ is the trace of the extrinsic curvature of the boundary intersecting with the spacelike surface of constant $t$ and ${\bf {\epsilon}}$ is the volume three-form (see {\cite{ccm} for details and notations). (\ref{CI}) implies that the quasilocal mass is given by \cite{ccm}
\begin{equation}
{\cal M}(r) = \frac{N}{2}[2rD(\phi)f_0-(r^2D(\phi))^{'}f] \period \label{mass1}
\end{equation}
Using this in both Einstein and string frames, with $f_0=1$, it is easy to check that
\begin{equation}
{\cal M}(\infty) = C =M_E. \label{m=C}
\end{equation} 
We call (\ref{CI}) as the conformally invariant (CI) prescription.  One interesting consequence of (\ref{CI}) is that every SSS conformally flat spacetime must have zero quasilocal mass. Note that an unique choice of a background prescription will imply an unique mass formula and vice versa.  For example, we first choose the HH prescription and uniquely get (\ref{mass}),  or we first demand conformal invariance and uniquely  get (\ref{CI}). It is also interesting to note that the result (\ref{eadm}) was recently obtained in \cite{ho} using a constant dilaton and Minkowski metric as the background. The authors identified $M_S$ as the energy instead of mass.  They interpreted (\ref{eadm}) as the difference between ADM mass and energy in string theory. They actually compared the mass due to the CI prescription with the mass due to HH prescription. Since $M_S$ is the conserved charge based on HH prescription, we interpret it as mass rather  than  energy. Our definition is that: !
For energy we mean the thermodynam
ic internal energy given by (\ref{energy}) which appears in the first law. For a conserved charge with respect to the timelike killing vector field, we call it mass. Based on the discussions given in \cite{ccm}, it is easy to check that the first law will not be satisfied in the string frame if one treats $M_S$ as the internal energy, that is, $\frac{\partial M_S}{\partial S}\neq T_H=\frac{1}{8\pi M_E}$. 
However, it was shown in \cite{ccm} that $\frac{\partial E}{\partial S} = T$ quasilocally and $T(\infty)=T_H=\frac{1}{8\pi M_E}$, no matter one use CI or HH prescription, in Einstein or string frame. Similar discussions apply to the electric case. 

In general relativity, that is, $D(\phi)=1$, both (\ref{mass}) and (\ref{mass1}) in HH and CI prescriptions are the same. However, when dealing with string theory or scalar-tensor gravity,  one has to decide which mass formulae to use.  We prefer HH prescription for three reasons.
First, $S_0$ in the CI prescription is too ``artificial''.  Imagine that only string theory (\ref{saction}) is known in our world, and we do not know what a conformal transformation is.  If we want to calculate the quasilocal and asymptotic mass for a solution, then how do we  know (\ref{CI}) is the background to choose?  Does that mean we may never be able to calculate mass from the string action and metric, if we do not know what Einstein metric and action are. 
Why do we need the knowledge of conformal transformation in order to calculate mass? On the other hand, in HH prescription, $S_0$ is naturally assumed to has same functional form of $S$. No knowledge of conformal transformations is required. Second, (\ref{mass1}) may yield an asymptotic mass which is inconsistent with Hamiltonian analysis, in an asymptotically non-flat spacetime unless $D(\phi)=1$. Take the $2+1$ conformal solution obtained in \cite{mz} as an example. The action and solution are:
\begin{equation}
S=\int  d^3x {\sqrt{-g}} [\left(1-{1\over 8}\Psi^2\right){\cal R}-(\nabla\Psi)^2
+ 2\Lambda],  \label{ctheory} 
\end{equation}
\begin{equation}
ds^2 = -F(r)dt^2 + {dr^2\over F(r)}+r^2d\theta^2, \quad F={\Lambda}\left[r^2-{M\over \Lambda}-{2\over r}\left({M\over 3\Lambda}\right)^{{3\over 2}}\right],  \quad  
\Psi=\sqrt{8}\sqrt{{\sqrt{{M\over 3\Lambda}}\over r+\sqrt{{M\over 3\Lambda}}}}.
\label{cbh}
\end{equation} 
This solution is asymptotically anti-de Sitter and therefore one expect that it should have a finite mass\footnote{The 2+1 BTZ black hole is asymptotically anti-De Sitter and admits a finite asymptotic mass.}. Both methods, the Hamiltonian method adapted in \cite{mz} and the $(2+1)$-dimensional  version of (\ref{mass}) in HH prescription with $M=0$ as the background derived in \cite{kevin}, consistently yield $M$ as the finite mass parameter identified at spatial infinity. 
However, it was shown in \cite{ccm} that the $(2+1)$-dimensional version of (\ref{mass1}) yields a diverging mass at spatial infinity. Thus 
(\ref{mass1}) yields a result which is not consistent with the result from the  method of Hamiltonian. In fact, one can roughly see that  (\ref{mass1}) has a tendency to yield a divergence in  asymptotic mass in an asymptotically non-flat spacetime. In (\ref{mass}), the divergence in $f_0$ can balance the divergence in $f$, and consequently $f_0-f$ can be finite. On the other hand, no such a ``symmetric'' expression exists in (\ref{mass1}).  Such a ``symmetry'' in (\ref{mass}) in HH prescription is expected since the background action has the same form as the rest of the  action.  

\section{Active, passive and inertial mass}

There is one more reason which makes us to prefer (\ref{mass}) over (\ref{mass1}). 
First we need to review the concepts of  active gravitational mass, passive gravitational mass and inertial mass in Newtonian Mechanics. A body experiences a force proportional to its passive gravitational mass (or passive mass, $m_p$) when put inside a background gravitational field. The same body produces a gravitational field which is asymptotically proportional to its active gravitational mass (or active mass, $m_a$). On the other hand,  the body undergoes an acceleration which is inversely proportional to its inertial mass ($m_i$).  The inertial mass can be regarded as the total contributions of mass of all sort of the body. In terms of  1-dimensional motion, we can summarize the above masses in Netwonian mechanics when the motion of body 1 is under the influence of body 2:
\begin{equation}
\frac{m_{1p}m_{2a}}{x^2}=m_{1i} {\ddot x}. \label{newt}
\end{equation}
In Newtonian mechanics, we assume $m_p=m_a=m_i$. 

In general relativity, one can also define  $m_p$, $m_a$ and $m_i$ for an asymptotically flat, and SSS spacetime. For example, $m_a$ of a SSS solution is obtained by comparing the weak field limit of the geodesic equation of a non-relativistic test particle with Netwon's second law \cite{wald}. It is given by
\begin{equation}
m_a =  \lim_{r \rightarrow \infty} \left(\frac{r}{2}\right)(g_{tt}+1) \period \label{m_a}
\end{equation}
The inertial mass can be calculated  from the radial part of the SSS metric by using gravitational stress-energy pseudo-tensor \cite{ran, myers}: 
\begin{equation}
m_i =  \lim_{r \rightarrow \infty} \left(\frac{r}{2}\right)(1-g^{rr})  \period \label{m_i}
\end{equation}
$m_p$ is set equal to $m_i$ due to the Weak Equivalent Principle (WEP).
Alternatively, $m_i$ is just the asymptotic mass calculated from (\ref{mass}) or (\ref{mass1}), or called the ADM mass, since it contains the total contributions of all sorts of mass of the solution. It has not yet known whether the active and passive masses are  equivalent in general relativity \cite{ran}. 

In scalar-tensor or string gravity (\ref{saction}), both passive and active mass are similarly defined as in general relativity \cite{ran}. Thus (\ref{m_a}) and (\ref{m_i}) are still valid  with string metric. However, we do not know which of (\ref{mass}) or (\ref{mass1}) at spatial infinity will give us the required inertial mass in string frame. We can use GHS solutions as examples.  Now, for the magnetic black hole in the string frame, we have:
\begin{equation}
m_a = M_E - \frac{Q_m^2}{2M_E}, \quad m_i = m_p = M_E+\frac{Q_m^2}{2M_E}. \label{mag_m}
\end{equation}
Similarly for the electric black hole in the string frame, we have:
\begin{equation}
m_a = M_E+\frac{Q_e^2}{2M_E}, \quad m_i = m_p = M_E - \frac{Q_e^2}{2M_E}. \label{elec_m}
\end{equation}
Finally for the Einstein metric, we have:
\begin{equation}
m_a = m_p = m_i = M_E.  \label{ein_m}
\end{equation}
Comparing  (\ref{mag_m}) and (\ref{elec_m}) with (\ref{madm}) and (\ref{eadm}), we see that the asymptotic mass calculated from (\ref{mass}), instead of (\ref{mass1}),  yields the inertial mass.  In fact, it is not hard to see that as $r\rightarrow\infty$, the mass formula from the HH prescription,  the equation (\ref{mass}), becomes the inertial mass (\ref{m_a}). It is the third reason that makes us to prefer (\ref{mass}) over (\ref{mass1}). Note that  in the string frame, the extreme condition implies that $m_a=0$ in the magnetic case (\ref{mag_m}) and $m_i=0$ in the electric case (\ref{elec_m}) respectively.  Different notions of mass are summarized in the following table: 

\vspace{0.3cm}

\begin{tabular}{lllll} \hline
{Frame}&{ CI Prescription}&{HH Prescription}&{Inertial Mass}&{Active Mass}\\ \hline

\vspace{0.1cm}

Einstein & $M_E$ & $M_E$ & $M_E$ & $M_E$ \\

\vspace{0.2cm}

String (mag) & $M_E$ &  $M_E + \frac{Q_m^2}{2M_E}$  & $M_E + \frac{Q_m^2}{2M_E}$ &  $M_E - \frac{Q_m^2}{2M_E}$   \\

\vspace{0.2cm}

String (elec) & $M_E$ & $M_E -\frac{Q_e^2}{2M_E}$  & $M_E -\frac{Q_e^2}{2M_E}$  & $M_E+\frac{Q_e^2}{2M_E}$ \\

\hline
\end{tabular}
\bigskip
\bigskip

It is worthwhile to mention that in the magnetic case, the difference 
in asymptotic mass between the Einstein metric and
string metric was previously  written down in equation (2.4) of \cite{cm}, 
but the method to obtain the mass is not a proper method: The authors in \cite{cm}
considered the asymptotic form of  $g_{tt}$ of the string metric in $r\rightarrow\infty$ limit, and identified the mass as the coefficient of the $\frac{1}{r}$ term. However, as we have discussed above, the significance of the asymptotic form of  $g_{tt}$     
is to determine the active gravitational mass rather than the asymptotic mass given in (\ref{mass}) or (\ref{mass1}) at spatial infinity. Thus their
original equation (2.4) is simply expressing the difference in the active gravitational mass in the Einstein and string metrics.  This method was also used in  \cite{xz} to identify the mass in the asymptotically flat Bekenstein black hole solution. The action is in the form of (\ref{E1}) with $D(\phi)=1-\frac{1}{6}\phi^2$. 
The Bekenstein solution reads
\begin{equation}
ds^2=-f^2(r)dt^2+\frac{dr^2}{f^2(r)}+r^2d\Omega^2 \comma \quad  f=\left(1-\frac{r_o}{r}\right) \quad  \phi=\frac{r_o}{r-r_o} \label{bek}
\end{equation}
The authors identified the mass by just using the asymptotic behavior of $-g_{tt}=f^2(r)$ and got $M=r_o$. In fact, what they got is the active gravitational mass. However, using  (\ref{bek}) in (\ref{mass}) or (\ref{mass1}),  it is easy to checked that the mass is also $r_o$. The reason is that the term $\frac{2D(\phi)}{r}+D'(\phi)=\frac{2}{r}\left(1+\frac{r_o^3}{6(r-r_o)^3}\right)$ falls off faster than the $f(r)$ so it becomes insignificant at spatial infinity, and as a consequence one can just identify the mass from the asymptotic $f(r)$ alone. 
By coincidence they used a wrong method but got a right answer on the mass. Generally speaking,  in scalar-tensor or string gravity, one cannot naively identify the mass parameter by using the asymptotic form of $g_{tt}$ of an asymptotically flat spacetime. Strictly speaking one should use (\ref{mass}) or (\ref{mass1}) to carry out the calculation explicitly, for any SSS spacetime with any asymptotic condition.  

\section{Summary}

To sum up, conformal transformations like (\ref{ct}) can induce
a difference in the asymptotic mass parameters in the two conformally related SSS spacetimes. Here we illustrate this by using the conformally related string and Einstein black holes in \cite{horowitz, ghs}. The reason is that the background action in the HH prescription is not conformally invariant. 
One can construct a conformally invariant background action but its definition is too artificial, it yields a result which is not consistent with Hamiltonian method of calculating mass, and it does not agree with the definition of inertial mass.  Finally we briefly mention some improper methods of identifying the mass parameter in SSS solution.

\end{document}